# Common Artifacts in Volume Rendering

Daniel Ruijters[1,2]

[1] Philips Healthcare, Image Guided Therapy Systems Innovation, Veenpluis 6, 5684PC, the Netherlands
[2] Technische Universiteit Eindhoven, Dept. Electrical Engineering, Den Dolech 2, 5612AZ Eindhoven

*Direct Volume Rendering is a popular and powerful visualization method for voxel data and other volumetric scalar data sets. Particularly, in medical applications volume rendering is very commonly used, and has become one of the state of the art methods for 3D visualization of medical data. In this article, some of the most common artifacts encountered will be discussed, and their possible remedies.*

## Introduction

Many medical imaging modalities have the capability to produce volumetric voxel data, such as CT, MR, ultrasound, etc. To fully appreciate the 3D shapes and morphologies within those data, volumetric render techniques have been developed. Visualization through volume rendering finds a wide range of applications in the medical domain, such as diagnostics [1], interventional imaging [2,3,4], stereoscopic rendering [5,6], etc. Since volume rendering is very commonly used for medical applications, it is important to understand which artifacts may arise, how to recognize them, what their root cause is, and which strategies can be followed to avoid or reduce them. This article will focus on the artifacts that are directly related to the volume rendering process. It will not deal with artifacts that are inherently present in the data, such as streak artifacts in CT data or partial volume effects in the voxel content [7]. It also important to note that this article focuses on the volume rendering equation above without any additions such as sophisticated lighting, shadows, etc.

## Direct Volume Rendering

Direct Volume Rendering is based on following the traversal of rays of light through the data. This is achieved by evaluating the rendering equation as introduced Kajiya [8] along the rays:

$$(1) \quad i = \int_0^\infty c(\vec{x}(t)) \cdot e^{-\int_0^x \tau(\vec{x}(t'))dt'} dt$$

Whereby $\vec{x}(t)$ represents a ray in 3D Euclidian space parametrized by $t$, and $i$ represents the resulting color of the ray, $c(\vec{x})$ is the emitted color at location $\vec{x}$, and $\tau(\vec{x})$ the light absorption at a particular location. Typically, $c$ and $\tau$ are determined by first interpolating the voxel data at location $\vec{x}$, which delivers a scalar value $v(\vec{x})$. Then, a transfer function is applied, often implemented as a lookup table, delivering $c(v(\vec{x}))$ as a RGB value and $\tau(v(\vec{x}))$ as a scalar value in the range [0,1].

While there are several volume rendering algorithms, they typically perform an approximate evaluation of Equation 1 by executing a discretized calculation of the integral, which can be written as follows:

$$\text{(2)} \quad i = \sum_{\{n=0\}}^{N} \left( \alpha_n c_n \cdot \prod_{\{n'=0\}}^{n} (1 - \alpha_{n'}) \right)$$

Whereby the ray is divided into N segments, and $\alpha_n$ represents the opacity of the segment, and $c_n$ its color. The segments are equidistant for many implementations, but this is not a hard requirement. The calculation of the segment colors and opacity may be as simple a single $c(\vec{x})$ and $\tau(\vec{x})$ lookup per segment, or may involve a more sophisticated evaluation.

The vast majority of the volume rendering algorithms do require a transition from the discrete voxel samples into a continuous expression in the 3D Euclidian space, meaning that the color and absorption for an arbitrary location $\vec{x}$ can be determined. This is typically accomplished through interpolation. Tri-linear interpolation is the most used interpolation scheme, motivated by its simplicity and hardware support.

## Artifacts

In this section, the most common artifacts will be presented. Also, their root cause and potential ways to overcome or reduce them will be discussed.

### Onion rings

The approximate discretization of Equation 1 into Equation 2 may lead to sampling artifacts which can manifest themselves as so called 'onion rings' as depicted in Figure 1(a). There are several factors that influence the severity of this artifact:

- Sample distance (or sample frequency): When the segment length for Equation 2 is reduced (which is equivalent to increasing the number N of samples), the artifacts will be reduced, as the discretization error becomes smaller. It should be noted that when the segment length is reduced from $d_1$ to $d_2$, the segment opacity should be adjusted according to:

$$\text{(3)} \quad \alpha_2 = 1 - (1 - \alpha_1)^{\frac{d_2}{d_1}}$$

in order to account for the influence of the reduced length to the opacity in the product series in Equation 2. This is illustrated in Figure 2. In order to optimize for speed while yielding few artifacts, Quatrin Campagnolo *et al.* [9] utilize an adaptive segment length, whereby smaller segments are used when the contribution to the ray color is large, and longer segments when the contribution is low.

- Equidistant sampling: The stripes in the 'onion ring' pattern arise from the fact that the rays in neighboring pixels are sampled at similar distances, yielding comparable errors (in magnitude and sign). When the sample locations of the rays in adjacent pixels is varied (e.g., by using variable sample distances, or a random offset at the beginning of each ray), the stripe pattern is broken. This leads to a more 'dithered' image, as the errors are still present but now randomly distributed over the pixels.

- Algorithm for the segment colors and opacity in Equation 2: In its simplest form these are determined by a simple interpolated sample at the segment center location in the voxel data set, after which the lookup table is applied. This procedure, however, yields the strongest artifacts, as can be seen in Figure 1(a). A more sophisticated approach has been presented by Engel *et al.* [10], which takes the interpolated voxel values $v_0(\vec{x}(t_0))$ and $v_1(\vec{x}(t_1))$ at the beginning and end of the segment. Then, a more accurate estimation of the integral in Equation 1 for the segment is performed, utilizing a pre-computed 2-dimensional lookup table. This method, called 'pre-integrated volume rendering', is illustrated in Figure 1(b). A more sophisticated approach to pre-integration, using three scalar samples per segment instead of two is presented by El Hajjar *et al.* [11], while Marchesin and de Verdière [12] describe a semi-analytical method and de Boer *et al.* [13] present a higher-order integration scheme. Csébfalvi provides an overview of pre-integration in volume rendering in [14].

- Transfer function or lookup table: The onion rings are most visible when there are abrupt transitions in material color and transparency. At such transitions, moving a sample point forward or backwards only a bit will lead to large differences in the value delivered by the transfer function. A smooth transfer function (or lookup table with only small differences between adjacent entries) will deliver far fewer 'onion ring' artifacts.

**Linear interpolation artifacts**

Linear interpolation is known to produce artifacts, also in other applications than volume rendering. Krylov *et al.* state in [15]: "Every linear resampling method has its own trade-off between three types of artifacts: ringing, aliasing and blur". In volume rendering the aliasing and blurring effects are most pronounced. Particularly, the aliasing translates in star-like shapes for very small object, such as one voxel wide vessels (see Figure 3).

There are three strategies to avoid or reduce the linear interpolation artifacts:
1. Create a higher resolution data set during the acquisition and reconstruction phase. Obviously, this is only possible when these phases can be controlled by the user. When the data set is simply a given, this is not possible. The downside of this approach is larger data storage requirements, larger memory footprint, and lower frame rates during rendering.
2. Create a higher resolution version of a given data set in a pre-processing step, using a more advanced interpolation scheme, such as cubic interpolation. The downsides are again larger memory footprint, and lower frame rates during rendering.
3. Replace tri-linear interpolation by a more advanced on-the-fly interpolation approach [16,17]. Often this is combined with a prefiltering step, whereby all voxel values are pre-processed, and then a combination of local lookups is performed to perform more accurate interpolations during the evaluation of the rays. An example is pre-filtered cubic B-spline interpolation [18]. The advantages of this approach are that it does not require larger data storage or memory footprint. It does come, however, at a performance penalty.

Still, tri-linear interpolation remains immensely popular in volume rendering applications. This can be contributed to its simplicity in implementation, its hardware support on among others GPUs [19], and the fact that the artifacts are spatially very small for high resolution data sets.

**Jerky edges**

The last artifact that is being discussed here are the jerky edges, as depicted in Figure 4(a). They occur when there the rays are sampled at a fixed distance, and there is an incomplete segment left at the boundary of the voxel volume, or at a cutting surface. The solution is to introduce a smaller segment that fills the distance from the last full segment to the volume boundary or cutting surface, as illustrated in Figure 4(b). Of course, it is important to adjust the opacity of the partial segment according to Equation 3.

## Interactive rendering

Many solutions or workarounds that reduce the artifacts' impact have a negative influence on the frame rates that can be achieved, as indicated in the previous section. Particularly, for clinically interventional applications as described in [3,4,5], however, require minimum frame rates and latency to accommodate hand-eye coordination. Strategies that balance the trade-off between artifact level and interactiveness may be required for such applications. These may sacrifice some image quality by allowing more artifacts when performance is needed (e.g., when manipulating the volume or scene), and render in the highest quality when possible (e.g., when the camera is not being moved). The following strategies can be applied in this respect:

- Increase the sample distance (i.e., segment length) when fast rendering is needed, and reduce it when high quality is demanded, see Figure 5.
- Use simple single sampling per segment for fast rendering, and pre-integrated rendering for higher quality, see Figure 5.
- Apply linear interpolation for fast rendering, and cubic interpolation for higher quality, see Figure 5.
- Use a lower resolution data (or mipmap pyramids [20,21]) for fast rendering, and high resolution for higher quality.
- Reduce the resolution of the rendered output view, leading to fewer rays, for fast rendering, and increase its resolution for higher quality. Table 1 presents the render times per frame depending on the output view resolution.

There are also many measures that can be employed to improve the frame rate that do not hamper the rendered image quality, such as empty space skipping, early ray termination, and GPU acceleration [19].

For interventional applications also precomputing times can be of relevance. The calculation of e.g., the coefficients for cubic B-spline interpolation can become significant. This procedure can be accelerated by employing the GPU [18]. Table 2 presents the results of pre-computing these coefficients for different hardware platforms and implementations.

## Conclusions

In this article, the most common and prominent artifacts that occur in direct volume rendering have been discussed. The root causes have been pinpointed and solutions and workarounds to avoid or reduce these artifacts have been identified. Unfortunately, the solutions typically come at a computational performance penalty, which is also visible in the experimental results. While hardware acceleration and smart techniques, such as adaptive sample distances and multi-resolution architectures, can be employed to reduce the negative impact on the computational performance, a trade-off between fast rendering and high image quality is often unavoidable. Still, in cases where computation time is less of a bottle neck (e.g., because the scene and camera position is not changing), it is possible to create volume rendered images with little to no artifacts.

## References

[1] S. Perandini, N. Faccioli, A. Zaccarella, T.J. Re, R.P. Mucelli, "The diagnostic contribution of CT volumetric rendering techniques in routine practice", The Indian journal of radiology & imaging, 20(2):92-97, 2010. https://dx.doi.org/10.4103/0971-3026.63043

[2] L. Spelle, D. Ruijters, D. Babic, R. Homan, P. Mielekamp, J. Guillermic, and J. Moret, "First clinical experience in applying XperGuide in embolization of jugular paragangliomas by direct intratumoral puncture", International Journal of Computer Assisted Radiology and Surgery 4(6):527-533, November 2009. https://doi.org/10.1007/s11548-009-0370-6

[3] J.A. Garcia, S. Bhakta, J. Kay, K.-C. Chan, O. Wink, D. Ruijters, and J.D. Carroll, "On-line multi-slice computed tomography interactive overlay with conventional X-ray: A new and advanced imaging fusion concept", International Journal of Cardiology 133(3):e101-e105, April 17, 2009. https://doi.org/10.1016/j.ijcard.2007.11.049

[4] P. Ambrosini, I. Smal, D. Ruijters, W. Niessen, A. Moelker, and T. van Walsum, "A Hidden Markov Model for 3D Catheter Tip Tracking with 2D X-ray Catheterization Sequence and 3D Rotational Angiography", IEEE Transactions on Medical Imaging 36(3):757-768, March 2017. https://doi.org/10.1109/TMI.2016.2625811

[5] D. Ruijters and S. Zinger, "IGLANCE: Transmission to Medical High Definition Autostereoscopic Displays", in Proceedings of the 3DTV-CONFERENCE 2009: The True Vision - Capture, Transmission and Display of 3D Video, May 4-6, 2009, Potsdam (Germany), 4 pages. https://doi.org/10.1109/3DTV.2009.5069626

[6] S. Zinger, D. Ruijters, L. Do, and P.H.N. de With, "View Interpolation for Medical Images on Autostereoscopic Displays", IEEE Transactions on Circuits and Systems for Video Technology 22 (1):128-13, 2012. http://dx.doi.org/10.1109/TCSVT.2011.2158362

[7] S.L. Wood, and S. Napel, "Artifacts and illusions in surface and volume rendering", in Proceedings of the 14th Annual International Conference of the IEEE Engineering in Medicine and Biology Society, 1992. https://doi.org/10.1109/IEMBS.1992.5762182

[8] J.T. Kajiya. "The rendering equation", Computer Graphics, in Proceedings of SIGGRAPH'86, 20(4):143–150, 1986. https://doi.org/10.1145/15886.15902

[9] L. Quatrin Campagnolo, W. Celes, and L.H. de Figueiredo, "Accurate Volume Rendering based on Adaptive Numerical Integration", in Proceedings of the 28th SIBGRAPI Conference on Graphics, Patterns and Images, pp. 17-24, 2015. https://doi.org/10.1109/SIBGRAPI.2015.27

[10] K. Engel, M. Kraus, and T. Ertl, "High-quality Pre-integrated Volume Rendering using Hardware-Accelerated Pixel Shading", in Proceedings of the 2001 Eurographics workshop on Graphics hardware, pp. 9-16, 2001. https://dx.doi.org/10.1145/383507.383515

[11] J.-F. El Hajjar, S. Marchesin, J.-M. Dischler, and C. Mongenet, "Second Order Pre-Integrated Volume Rendering", in Proceedings of the IEEE Pacific Visualization Symposium, pp. 9-16, 2008. https://doi.org/10.1109/PACIFICVIS.2008.4475453

[12] S. Marchesin and G. Colin de Verdière, "High-Quality, Semi-Analytical Volume Rendering for AMR Data", IEEE Transactions on Visualization and Computer Graphics 15(6):1611-1618, 2009. https://doi.org/10.1109/TVCG.2009.149

[13] M. de Boer, A. Gröpl, J. Hesser, and R. Männer, "Reducing Artifacts in Volume Rendering by Higher Order Integration", in Proceedings of Late Breaking Hot Topics, Visualization 97, pp. 1-4. 1997.

[14] B. Csébfalvi, "An Evaluation of Prefiltered Reconstruction Schemes for Volume Rendering", IEEE Transactions on Visualization and Computer Graphics 14(2):289-301, 2008. https://doi.org/10.1109/TVCG.2007.70414

[15] A. Krylov, A.V. Nasonov, and A. Chernomorets, "Combined linear resampling method with ringing control", in Proceedings of the 19th International Conference on Computer Graphics and Vision, GraphiCon, 2009.

[16] N. Gavrilov, and V. Turlapov, "Advanced GPU-based Ray Casting for Bricked Datasets", in Proceedings of SIGGRAPH 2012, Los Angeles, California, August 5-9, 2012. https://doi.org/10.1145/2342896.2343038

[17] N. Gavrilov, and V. Turlapov, "Volume Ray Casting quality estimation in terms of Peak Signal-to-Noise Ratio", in Proceedings of EUROGRAPHICS, 2013. http://dx.doi.org/10.2312/conf/EG2013/posters/007-008

[18] D. Ruijters, and P. Thévenaz, "GPU Prefilter for Accurate Cubic B-Spline Interpolation", The Computer Journal 55(1):15-20, January 2012. https://doi.org/10.1093/comjnl/bxq086

[19] D. Ruijters, and A. Vilanova, "Optimizing GPU Volume Rendering", Journal of WSCG 14(1-3): 9-16, January 2006.

[20] S. Guthe and W. Strasser, "Advanced Techniques for High Quality Multiresolution Volume Rendering", In Computers & Graphics, pp. 51-58, 2004. https://doi.org/10.1016/j.cag.2003.10.018

[21] R. Carmona, G. Rodríguez, and B. Fröhlich, "Reducing Artifacts between Adjacent Bricks in Multi-resolution Volume Rendering", in Proceedings of Advances in Visual Computing, 5th International Symposium, ISVC 2009, Las Vegas, NV, USA, November 30 - December 2, 2009. http://dx.doi.org/10.1007/978-3-642-10331-5_60

# Figures, Tables

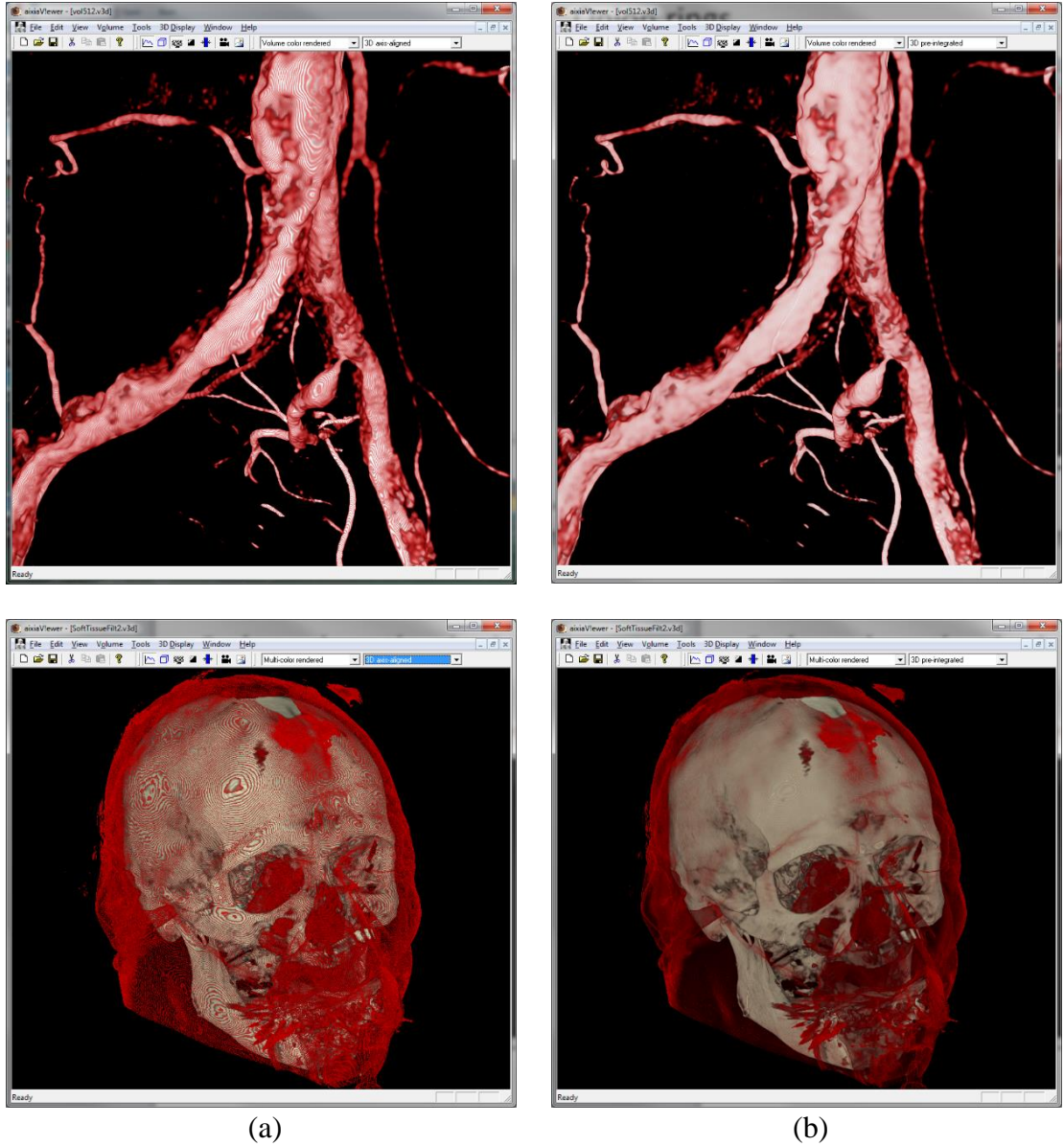

(a)                                                          (b)

*Figure 1: Examples of 'onion ring' artifacts. In column (a) each segment color and opacity is determined by a single sample. In column (b) the segments are determined by a pre-integrated lookup table. While the 'onion ring' artifacts are much more severe in column (a), they are not completely gone in column (b), as the pre-integrated approach still suffers from a small error.*

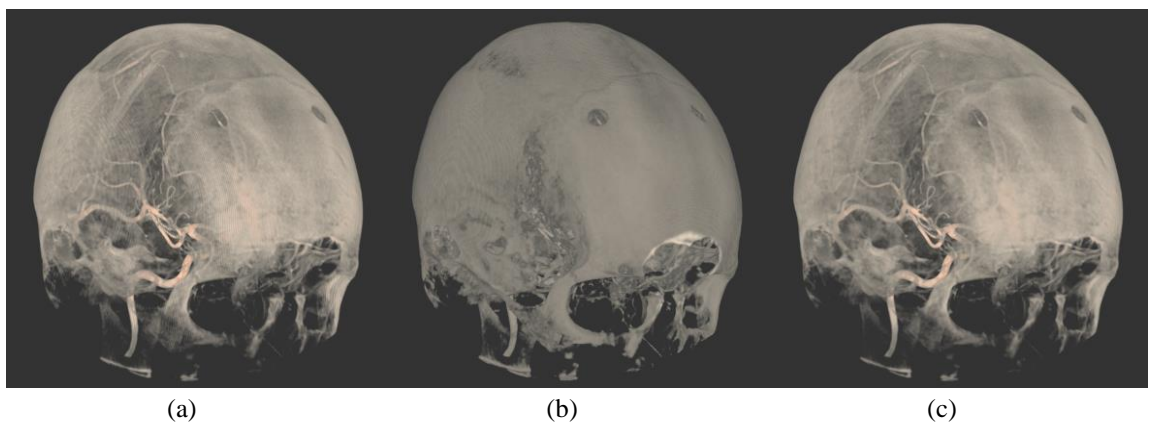

*Figure 2: (a) Each ray is rendered with with 1 sample per voxel. Some onion ring artifacts are visible. (b) The sample rate is set to 10 samples per voxel, but the opacity is not corrected for the change sample segment length, which results in an overall reduced translucency. (c) Again the sample rate is set to 10 samples per voxel, but now the opacity is corrected. The appearance is similar to the one in (a), but the onion ring artifacts are severely reduced.*

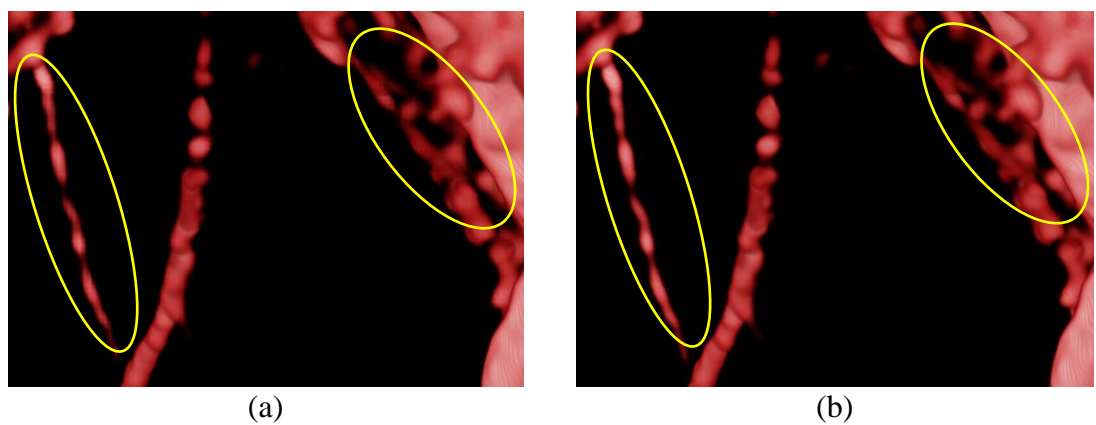

*Figure 3: (a) Volume Rendering of vessel fragment using linear interpolation. Yellow delineation shows most prominent star-shaped linear interpolation artifacts. (b) Same fragment using cubic interpolation [18]. Star shaped artifacts are not present anymore.*

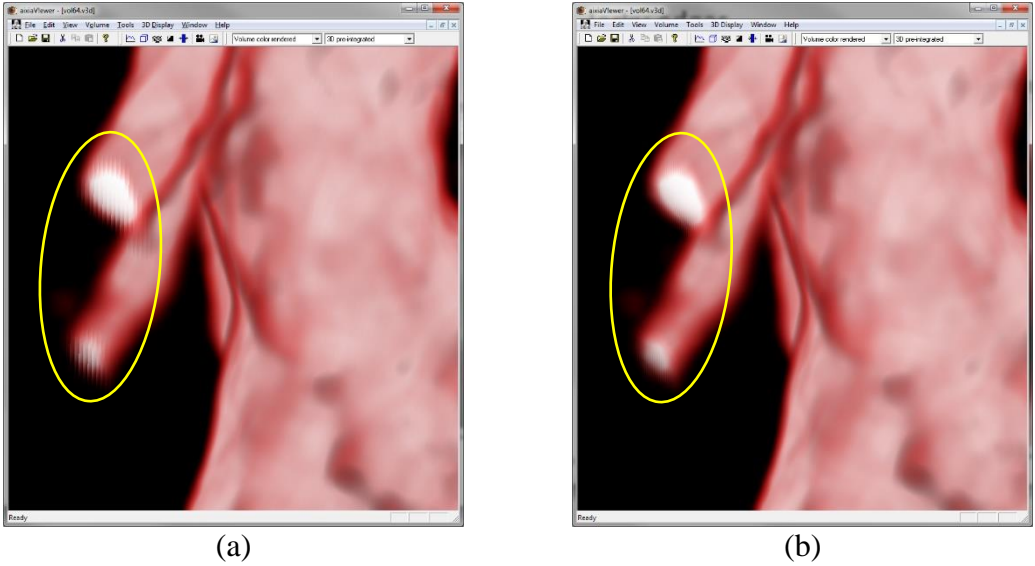

(a)                  (b)

*Figure 4: (a) Jerky edges at the edge of the volume. (b) Same fragment rendered with a higher sample frequency.*

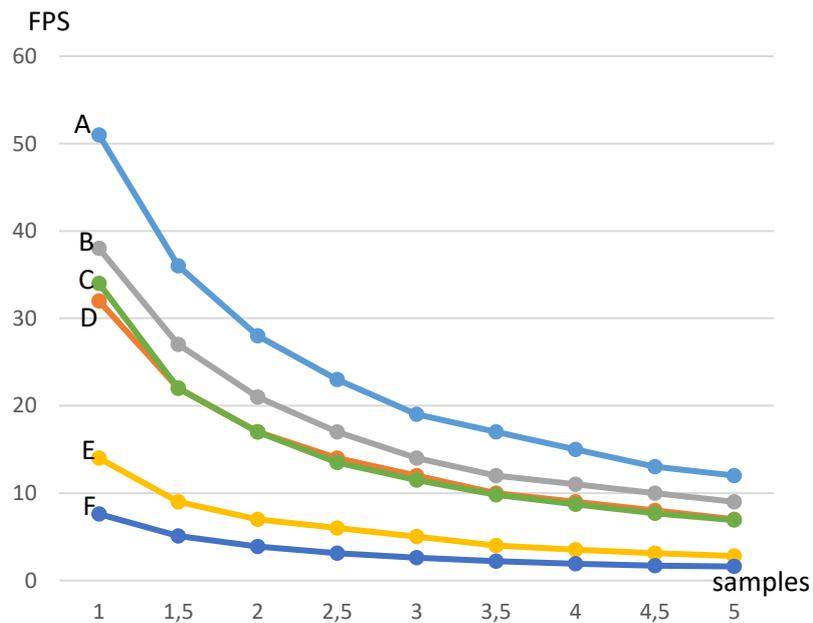

*Figure 5: Frame rate as function of samples per voxel (reciprocal of the distance between samples). A: simple volume rendering of a $256^3$ voxel dataset. B: pre-integrated rendering of the same dataset. C: pre-integrated rendering with Phong shading (advanced lighting) of the same dataset. D: simple volume rendering of a $512^3$ voxel dataset. E: same as A, but with tri-cubic interpolation instead of tri-linear interpolation. F: pre-integrated tri-cubic interpolated rendering of the $256^3$ voxel dataset. All measurements were performed on Intel Core i7 2.7 GHz machine with a nVidia Quadro 1000M, in a $1000^2$ viewport and no space-leaping or early ray-termination was used.*

| resolution   | pixels  | 4MB VR    | 200MB VR  |
|--------------|---------|-----------|-----------|
| 512 x 512    | 262144  | 31.67 ms  | 81.97 ms  |
| 800 x 600    | 480000  | 34.20 ms  | 114.63 ms |
| 1024 x 768   | 786432  | 50.83 ms  | 163.93 ms |
| 1280 x 1024  | 1310720 | 97.09 ms  | 218.34 ms |
| 1920 x 10809 | 2073600 | 114.94 ms | 280.90 ms |

*Table 1: Frame render times in ms for two different voxel data sets (of 4 MB and 200 MB) for different resolutions.*

| **CPU**    | Core i7 2.3 GHz     | Xeon E5 2.4 GHz     | Xeon E5 3.6 GHz     | Xeon E5 3.5 GHz         |
|------------|---------------------|---------------------|---------------------|-------------------------|
| **GPU**    | nVidia Quadro 1000M | nVidia Quadro 4000  | nVidia Quadro 4000  | nVidia GeForce GTX 1080 |
| **Simple** | 6.5 sec             | 9.1 sec             | 6.3 sec             | 7.3 sec                 |
| **OpenMP** | 1.6 sec             | 2.7 sec             | 1.6 sec             | 1.9 sec                 |
| **OpenCL** | 3.4 sec             | 1.6 sec             | 1.5 sec             | 1.0 sec                 |

*Table 2: Preprocessing step computation times for cubic B-spline interpolation coefficients, using different implementations (simple, OpenMP, OpenCL) and hardware combinations.*